# The AEROS ocean observation mission and its CubeSat pathfinder

**Rute Santos[(1),(4)], Orfeu Bertolami[(2)], E. Castanho[(3)], P. Silva[(3)], Alexander Costa[(4)], André G. C. Guerra[(4)], Miguel Arantes[(4)], Miguel Martin[(4)], Paulo Figueiredo[(4)], Catarina M. Cecilio[(5)], Inês Castelão[(5)], L. Filipe Azevedo[(6)], João Faria[(6)], H. Silva[(7)], Jorge Fontes[(8)], Sophie Prendergast[(8)], Marcos Tieppo[(9)], Eduardo Pereira[(9)], Tiago Miranda[(9)], Tiago Hormigo[(10)], Kerri Cahoy[(11)], Christian Haughwout[(11)], Miles Lifson[(11)], Cadence Payne[(11)]**

*[(1)] Departamento de Geociências, Ambiente e Ordenamento do Território, Faculdade de Ciências, Universidade do Porto, Rua do Campo Alegre s/n, 4169-007 Porto, Portugal, up201403484@up.pt*

*[(2)] Departamento de Física e Astronomia, Faculdade de Ciências, Rua Do Campo Alegre, s/n 4169-007, Porto, Portugal, orfeu.bertolami@fc.up.pt*

*[(3)] AD AIR Centre, Terinov-Canada de Belém S/N, Terra Chã, 9700-702, Angra do Heroísmo, Ilha Terceira, Portugal, info@aircentre.org*

*[(4)] CEiiA, Av. Dom Afonso Henriques, 1825. 4450-017 Matosinhos, Portugal, (+351) 220 164 800, space@ceiia.com*

*[(5)] CoLAB +ATLANTIC, Edifício LACS, Estrada da Malveira da Serra 920, 2750-834 Cascais, Portugal, info@colabatlantic.com*

*[(6)] DSTelecom, Rua de Pitancinhos 4700-727 Braga, Portugal, (+351) 253 109 500, geral@dstelecom.pt*

*[(7)] EDISOFT S.A., Rua Calvet Magalhães 245, 2770-153 Paço de Arcos, Portugal, (+351) 212 945 900, info@edisoft.pt*

*[(8)] Institute of Marine Sciences - Okeanos, University of the Azores, Rua Professor Doutor Frederico Machado 4, 9901-862 Horta, Portugal (+351) 292 200 400, okeanos.secretariado@uac.pt*

*[(9)] Institute for Sustainability and Innovation in Structural Engineering (ISISE), University of Minho, Azurém Campus, 4800-058, Guimarães, Portugal (+351) 253 510 200*

*[(10)] Spin.Works S.A., Rua de Fundões n.º151, 3700-121 São João da Madeira, Portugal, (+351) 256 001 949, info@spinworks.pt*

*[(11)] Massachusetts Institute of Technology, 77 Massachusetts Avenue, Cambridge, MA 02139, USA kcahoy@mit.edu*

## ABSTRACT

AEROS aims to develop a nanosatellite as a precursor of a future system of systems, which will include assets and capabilities of both new and existing platforms operating in the Ocean and Space, equipped with state-of-the-art sensors and technologies, all connected through a communication network linked to a data gathering, processing and dissemination system. This constellation leverages scientific and economic synergies emerging from *New Space* and the opportunities in prospecting, monitoring, and valuing the Ocean in a sustainable manner, addressing the demand for improved spatial, temporal, and spectral coverage in areas such as coastal ecosystems management and climate change assessment and mitigation. Currently, novel sensors and systems, including a miniaturized hyperspectral imager and a flexible software-defined communication system, are being developed and integrated into a new versatile satellite structure, supported by an innovative on-board software. Additional sensors, like the LoRaWAN protocol and a wider field of view RGB camera, are under study. To cope with data needs, a Data Analysis Centre, including a cloud-based data and telemetry dashboard and a back-end layer, to receive and process acquired and ingested data, is being implemented to provide tailored-to-use remote sensing products for a wide range of applications for private and institutional stakeholders.

# 1 INTRODUCTION

Oceans cover around 71% of the Earth's surface, playing a vital part in the Earth's system.

The oceans are one of the world's largest carbon sinks, absorbing approximately a quarter of the carbon dioxide ($CO_2$) that is released into our atmosphere [1]. Due to the continuous increase of these emissions, we have seen changes in seawater properties that are putting at risk many ecosystems that are essential for numerous marine organisms to survive, such as phytoplankton (marine plants responsible for maintaining oxygen and serve as food for many other marine organisms). This puts the entire food web at risk and, in turn, has a direct effect on fisheries, leading to considerable impacts on food production and human communities [2]. Climate change also has a negative impact on coastal areas, which are the most densely populated and economically active areas on earth [3].

Essential Ocean Variables (EOV), including ocean colour, Sea Surface Height (SSH), Sea Surface Salinity (SSS), and Sea Surface Temperature (SST), are key to understand its environment, as changes influence global climate and weather. Monitoring the ocean using drifters, buoys, gliders, and many other platforms have been the traditional methods used. However, conventional methods have some limitations [4]. Satellite remote sensing has thus played a significant role in oceanic and climate studies, offering an effective platform to estimate these EOV with high-resolution, both at a regional and global scale, mainly due to its synoptic coverage in space, regular revisit cycles and cost-effectiveness [5]–[8].

Although there are hundreds of small satellites currently orbiting the Earth, only a small percentage has been focused on the oceans [8]. The main objective of the AEROS project is to tackle this issue, monitoring both ocean and coastal areas, in particular ocean colour (to detect ocean fronts and distribution of marine megafauna locations) and sea surface salinity, using a hyperspectral imager aboard a 3U small satellite. The data obtained from the AEROS mission considers the development and improvement of object detection algorithms, which is also a valuable contribution to other ocean remote sensing studies. Apart from the scientific advancements, as a technology-oriented project, it will advance the technical know-how of Portuguese entities.

The AEROS project is developed by EDISOFT, CEiiA, Spin.Works, DSTelecom, University of Minho (UMinho), Faculty of Sciences - University of Porto (FCUP), University of Algarve (UAlgarve), Insituto Superior Técnico (IST) and Colab +Atlantic, in partnership with Institute of Marine Sciences - Okeanos (IMAR), AIR Centre and MIT, as part of the MIT Portugal programme.

# 2 AEROS USE CASES

The AEROS mission addresses the demand for improved spatial, temporal, and spectral Earth Observation (EO) coverage for scientific advances in areas such as coastal ecosystem management and climate change assessment and mitigation. In this context, three main use cases are being tackled to delineate the AEROS mission: (1) Effect of essential ocean variables and distribution of marine megafauna; (2) Fisheries and aquaculture management; and (3) Ecosystem-based management strategies and monitoring of Marine Protected Areas (MPAs).

## 2.1 Essential Ocean Variables and Distribution of Marine Megafauna

Understanding the environmental drivers of animal movement patterns and distributions is critical to support smart management and conservation strategies in a changing planet and increasing human pressure. Data from satellites has become essential to monitor and study species such as elasmobranchs, allowing animal's movements to be tracked over extensive and often inhospitable habitats largely inaccessible to humans [9]. Combined EO and animal tracking data, allows researchers to investigate the mechanisms that drive those patterns. Researchers have identified multiple biotic and abiotic parameters such as SST, SSH, primary productivity, SSS, mesoscale oceanographic fronts and eddies, oceanographic currents, dissolved oxygen, and bathymetry to correlate with elasmobranch distribution in a dynamic environment [10]–[13].

The contribution from established EO missions still lack, however, the required spatial resolution for site or organism-level analyses [14]. Small satellite constellation provides a unique opportunity to complement existing dataset from large missions with high resolution data for specific regions [15].

AEROS will collect EO data simultaneously with animal tracking data from animal-borne tags, which will enable researchers to examine the relationship between tagged animals and oceanographic features, such as mesoscale eddies/fronts and biotic factors such as primary productivity, at the local/regional scale to support smart management and conservation actions.

### 2.2 Fisheries and Aquaculture Management

EO data plays a key role in optimizing the activities carried out by the fishing industry, promoting a sustainable consumption of resources.

Global fish production is estimated to have reached about 179 million tons in 2018, of which 82.1 million tons came from aquaculture [16]. With the increasing investment in marine aquaculture since the last decade, the analysis of environmental parameters, such as SST, ocean currents, wave height, SSS, underwater irradiance, and water quality in terms of Suspended Particulate Matter (SPM), and phytoplankton is key to prevent environmental impacts and socio-economic conflicts [17].

The EOVs obtained through EO techniques in combination with *in situ* data and numerical models can lead to sustainable production, better fish health and reduced waste handling of biomass, which are often endangered by phenomena such as Harmful Algal Blooms (HABs). EO can highlight ocean areas at greatest risk of bloom development, identifying combinations of warm waters, sediment flows or pollution run-off that might promote their growth.

AEROS will provide important ocean variables in both the nearshore and offshore regions, through high-resolution data. This is paramount for the design and enforcement of fishing legislation, and strengthens the basis for its management, by providing more rationally based and informed decisions [18]. These objectives are directly link with “FAO Code of Conduct for Responsible Fisheries”, a voluntary agreement which sets out principles and international standards of behaviour for responsible practices, ensuring the effective conservation, management, and development of living aquatic resources, with due respect for the ecosystem and biodiversity.

### 2.3 Ecosystem-Based Management Strategies and Monitoring of Marine Protected Areas

Phenomena such as terrestrial run-off, eutrophication, and coastal upwelling event can affect MPAs and their management. These phenomena can have an affect even if they are happening externally, and thus a constant data feed from satellite is key to trigger and pinpoint an *in situ* data acquisition inside the MPA, aiming at monitoring their effects.

By providing data that can be used to detect and monitor HABs [19] and for assessments of terrestrial pollution through the analyses of SPM and Coloured Dissolved Organic Matter (CDOM) [20], typical components of coastal waters and important environmental indicators for MPAs, AEROS will be capable of delivering important information on this matter. High spatial resolution is also significant for smaller scale phenomena analysis in MPAs monitoring, such as river plumes in estuarine areas, especially in areas with high density of agriculture activity and urban zones, which often present problems regarding eutrophication [21].

# 3 AEROS MISSION DEVELOPMENT

## 3.1 Mission Objectives and Requirements

With the use cases established, a set of operation and mission objectives and requirements have been derived for AEROS, providing a common understanding regarding the operation of the first satellite, as well as the future constellation. The main objectives include: (1) the development and operation of a modular and versatile small satellite platform (bellow 6 kg, with a 3U form factor); (2) the development and test of a small satellite payload to monitor ocean colour, fronts and fauna (based on a HyperSpectral Imager (HSI)), and of a payload for collecting data from *in-situ* observation and monitoring systems (provided by a Software Defined Radio (SDR)); and, finally, (3) develop and test data processing algorithms, built on top of a Data Analysis Centre (DAC), to gather, store, process, analyse and disseminate added value data. All these objectives must be implemented in a tight schedule (less than 36 months) and within a fixed budget, respecting Portugal 2020 and MIT Portugal programme rules.

## 3.2 Concept of Operations and Orbit Analysis

The AEROS concept (Figure 1) involves the satellite collecting hyperspectral and *in-situ* data over the Region of Interest (ROI), and relaying it to one of the two ground stations, the primary at Boecillo or the secondary at Santa Maria. The ROI consists of the Portuguese Extended Continental Shelf (Figure 2). A full mission concept of operations, including launch, commissioning and end of life operations can be seen in Figure 4.

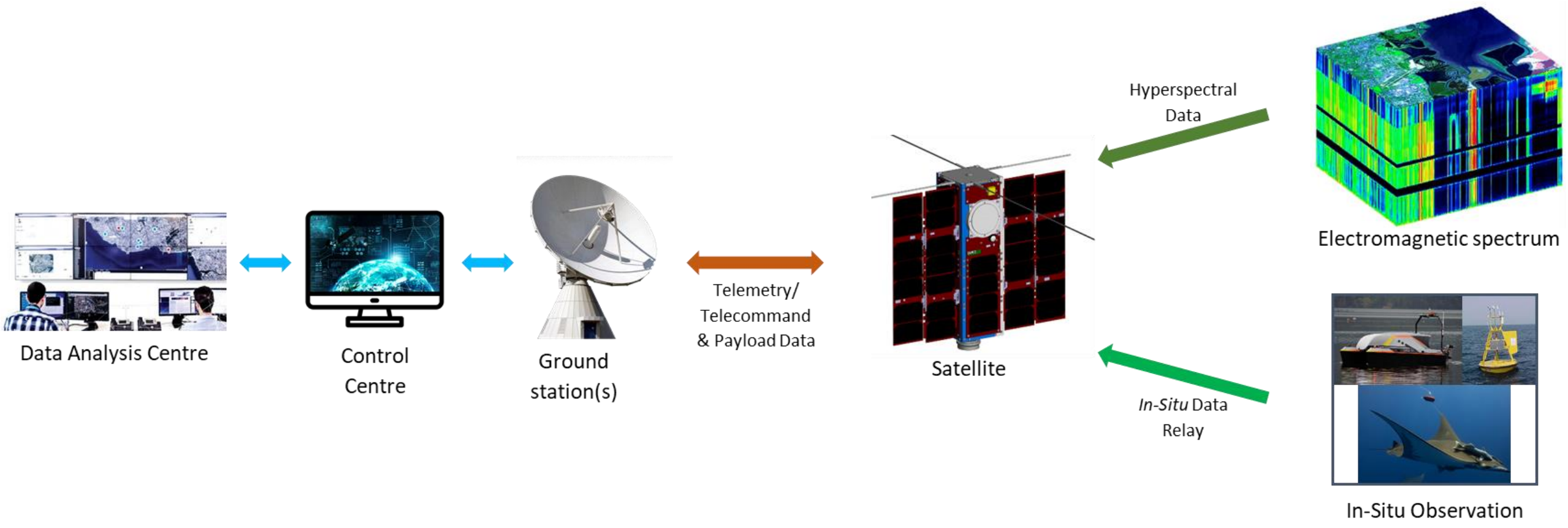

Figure 1. AEROS Schematic Preliminary Concept. Credit: CEiiA

The AEROS spacecraft will be launched in 2023 into a Sun-Synchronous Orbit (SSO), with a nominal deployment altitude of 500 km. Because the satellite does not feature propulsion, as the orbit decays it ceases to be SSO, and the ground-tracks "walk" over the ROI. Estimated orbital lifetimes for a variety of potential solar conditions are seen in Table 1. The nominal spacecraft area assumes it maintains a nadir-pointing configuration, with sun-tracking when possible and a 3U ram-area otherwise.

Table 1. AEROS Estimated Lifetime (lifetime < 1 year or >25 years highlighted in *red*)

| **Altitude [km]** | **Type** | **Drag Area [m$^2$]** | **Estimated Lifetime for Different Solar Activity Levels** | | | | | | |
|---|---|---|---|---|---|---|---|---|---|
| | | | *-3σ* | *-2σ* | *-1σ* | *0σ* | *+1σ* | *+2σ* | *+3σ* |
| 500 | Tumble, no panels | 0.0160 | 21.7 | 17.6 | 15.2 | 13.9 | 12.8 | 11.7 | 10.8 |
| 500 | Nominal | 0.0969 | 2.6 | 2.3 | 1.9 | 1.6 | 1.4 | 1.2 | *0.98* |
| 520 | Tumble, no panels | 0.0160 | *29.8* | *26.5* | 24.2 | 20.9 | 16.5 | 14.6 | 13.5 |
| 520 | Nominal | 0.0969 | 3.4 | 2.9 | 2.5 | 2.2 | 1.9 | 1.6 | 1.4 |

Previous studies included estimating power generation, orbital lifetime, and coverage of the ROI [22], based on initial assumptions, as the launch vehicle and precise orbital parameters were not available. Preliminary analysis performed now, with the latest orbit data, launcher (Falcon 9) and spacecraft configuration, indicates that these new parameters, versus earlier versions, do not impose significant issues for sensor coverage, communication link, or power generation (Figure 3).

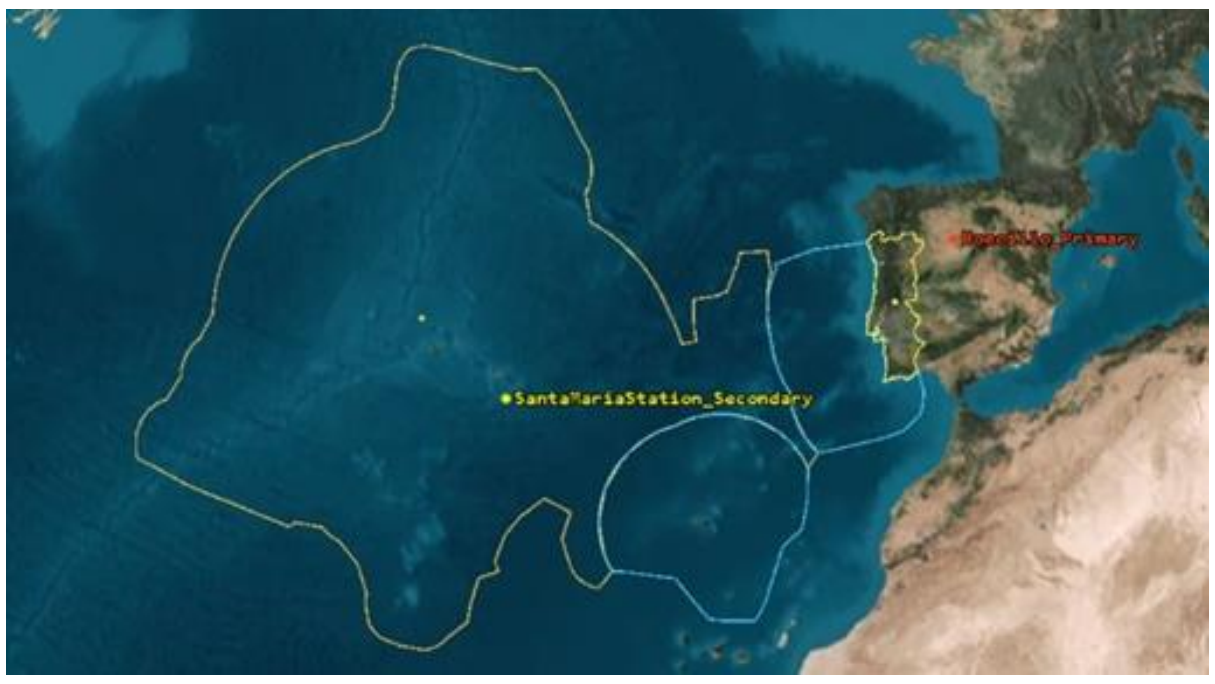

Figure 2. The AEROS ROI (i.e. the Proposed Extended Continental Shelf). Credit Miles Lifson

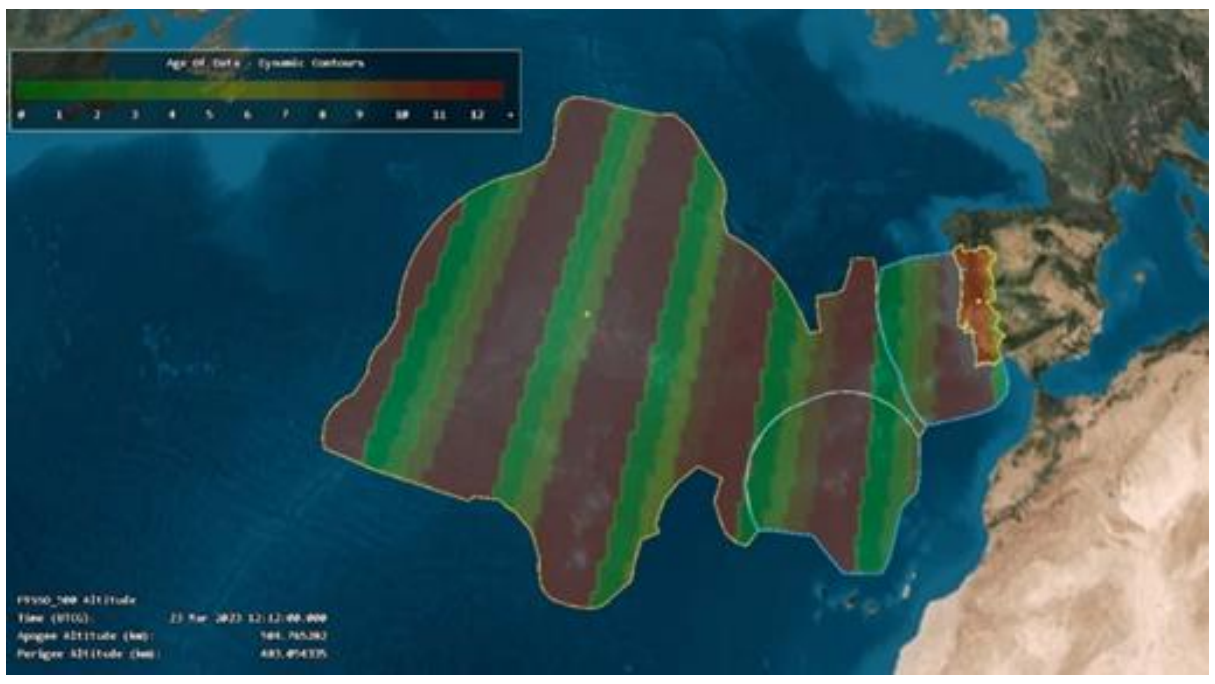

Figure 3. ROI coverage snapshot visualizing over at a particular moment. Credit Miles Lifson

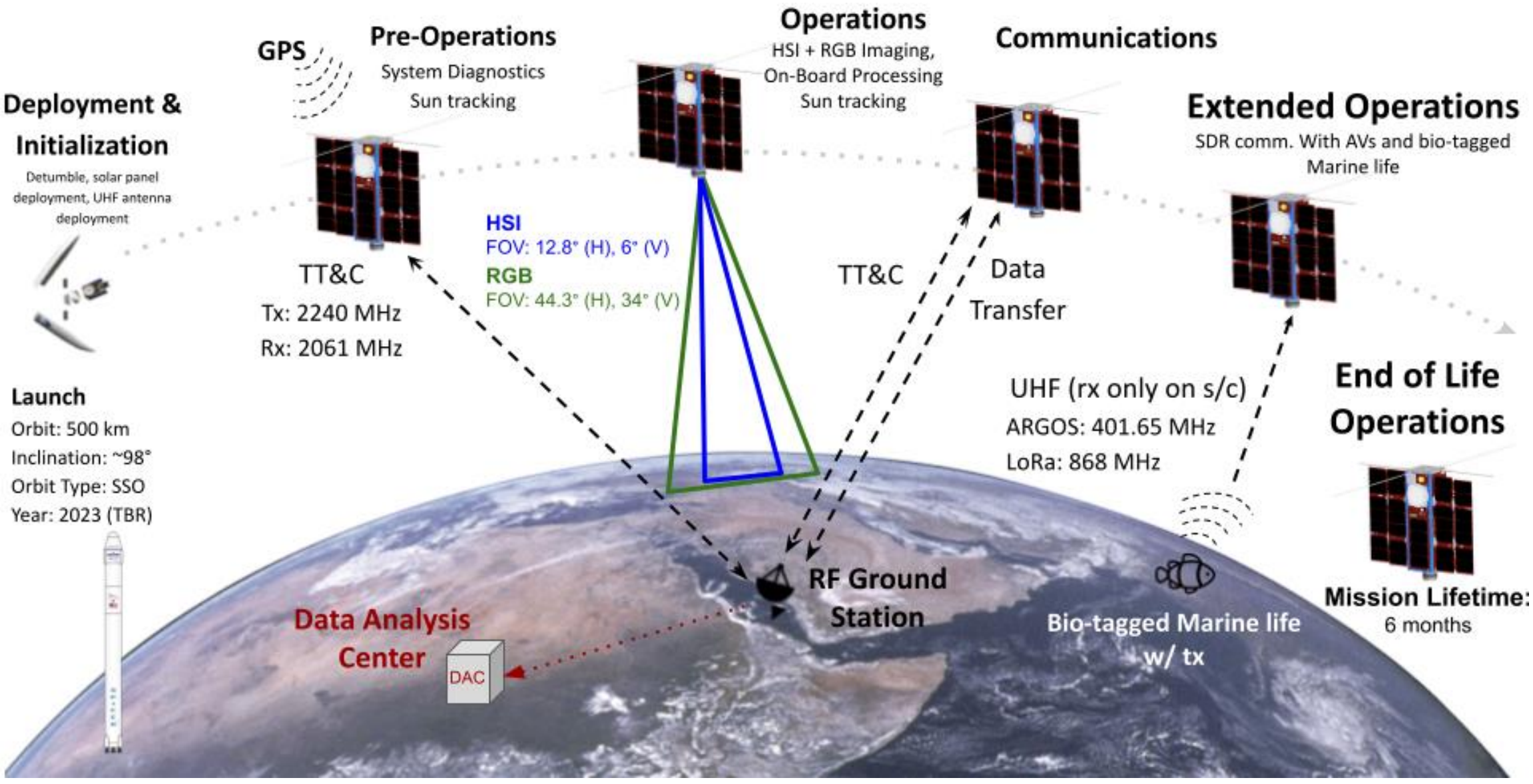

Figure 4. AEROS Mission Concept of Operations. Credit: Candence Payne

## 4 PAYLOADS

### 4.1 Hyperspectral Imager

There have been some in-orbit demonstrations of small hyperspectral cameras for small satellites (e.g. Cosine's Hyperscout, Dragonfly Mantis). Their multiple and narrower bands allow for a higher spectral resolution when compared with multispectral sensors, since it allows having more precise reflectance curves which can then be measured and compared, making it useful for environmental monitoring and complex oceanic environments [23].

For AEROS, based on its use cases, a miniaturized HSI has been considered for the primary payload. This HSI (Figure 5), being developed by Spin.Works, has 150 contiguous, 10 nm narrow, bands ranging between 470 nm and 1000 nm, covering the visual and near-infrared spectrum. The HSI package, based on an ams LS150 detector [24], has a mass of around 300 g with a camera focal length of 50 mm, a f/2.8 and a Field of View (FoV) of: 12.8º (full angle and horizontal), 6.0º (vertical), 14.6º (diagonal).

The HSI is a push-broom scanning device, where each camera row (or group of rows) is used to collect light in a narrow spectral band, while the camera columns are aligned with the orbital motion to ensure every spectral band is captured. The 2-megapixel sensor has a ground resolution of about 55 m, resulting on a swath of about 110 km. With 10-bit images, a single square-shaped hypercube contains about 3.2 Gbits. However, in practice, a small number of bands (10 to 15) will be selected to reduce the data storage requirements to less than 1/10 of the original size. The AEROS data processor operates in tandem with the imager, performing real-time 2-axis image alignment (trigger control and jitter compensation) at > 25 fps, being responsible for both hypercube reconstruction and lossless data compression (using the CCSDS 123.0-B-2 algorithm), reducing the overall data storage requirements by a factor >2.

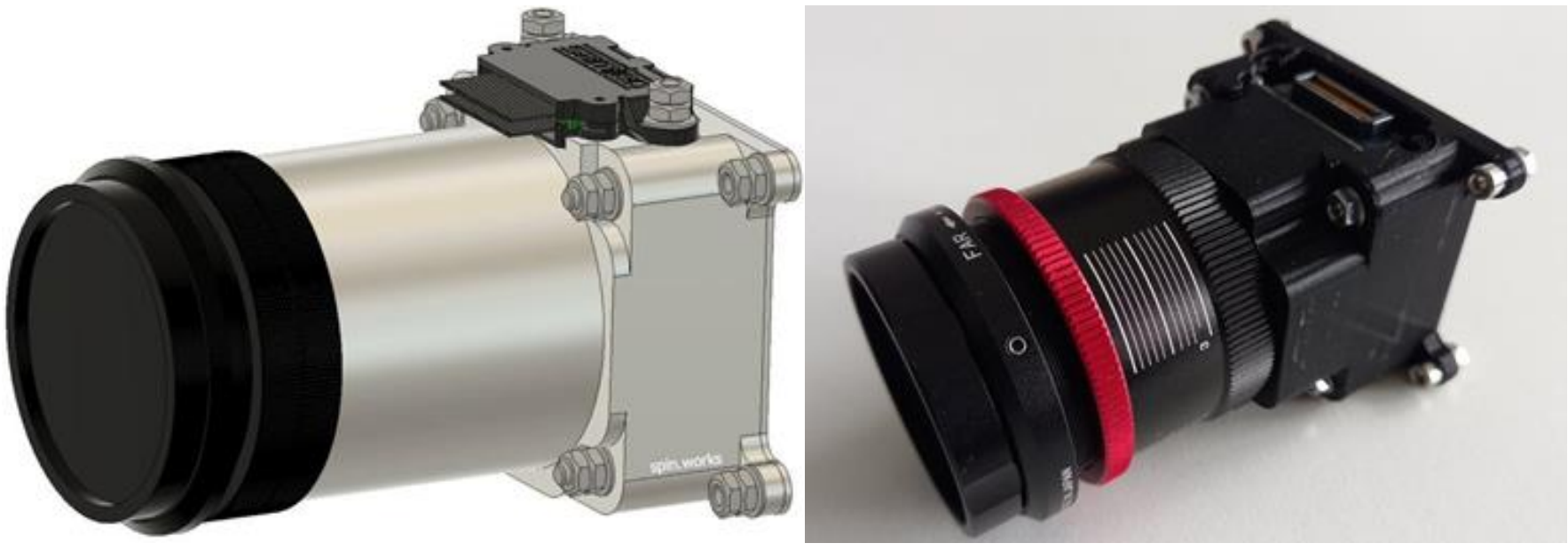

Figure 5. Hyperspectral Imager Design (left) and prototype (right). Credit: Spin.Works

## 4.2 Software Defined Radio

AEROS secondary payload, an SDR, will enable communication with ground assets (both bio-tagged marine life and autonomous vehicles) equipped with transmitter such as ARGOS and or LoRa.

SDR advantages include the reconfigurability of the Radio Frequency (RF), with tuning made through adjustment of variables in the code. This feature enables a new paradigm in space communication, with RF updates made from the ground [25].

The system in Figure 6 illustrates the SDR payload on-board the AEROS. The output of the SDR are two different files, the In-Phase and Quadrature (IQ) raw data and the result of demodulation which are bits extracted from the Phase-Shift Keying (PSK) signal. This approach ensures that any problem, like bit loss during demodulation, still guarantees access to the original signal.

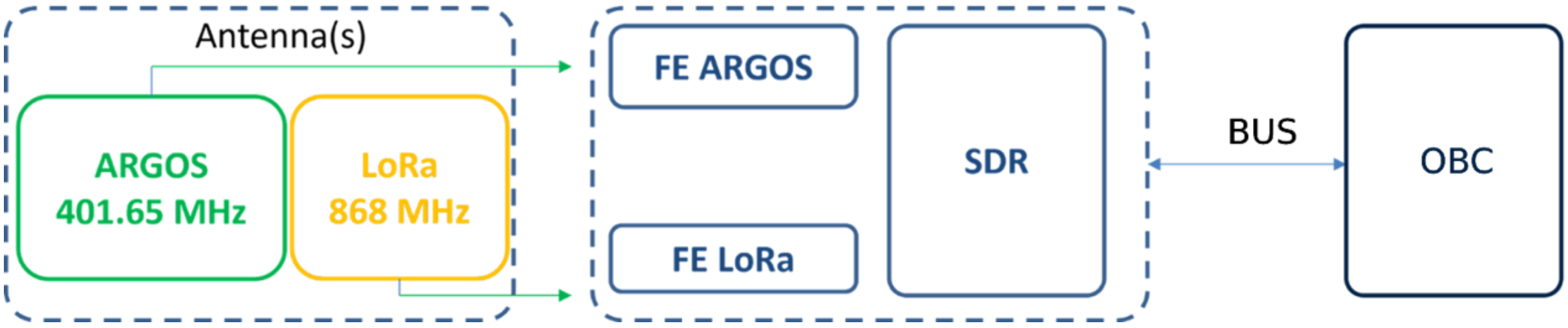

Figure 6. Architecture of the SDR. Credit: João Faria

### *4.2.1 ARGOS*

ARGOS is a unique worldwide location and data collection system dedicated to studying and protecting the environment. Although the ARGOS system has its own satellite constellation, providing world coverage, AEROS is seeking to build the capability of collecting transmitter data from its own assets. The objective is to demodulate the data on-board and then send both raw IQ files and the bits extracted from the PSK signal in a bin file to decrease the memory usage on-board of the payload.

### *4.2.2 LoRa & LoRaWAN*

Similar to ARGOS, the objective is to acquire the signal, transmitted by end devices, with the AEROS satellite and re-send it back to ground for post-processing. In this perspective, the satellite works as a gateway, collecting LoRa devices data. To ensure a proper acquisition, the system is designed to use in open fields with no obstacles in the line of sight between sensor and satellite (especially suited to ocean monitoring) LoRa devices have a low power consumption and uses globally available unlicensed bands, being easy to be integrated. It is important to distinguish LoRa, the physical layer of a spread spectrum modulation technique, from LoRaWAN, a low power, wide area networking standard.

### 4.3 RGB Camera

The AEROS satellite will feature a small RGB camera, the 5 MP Crystalspace CAM1U CubeSat Camera [26]. The primary purpose of the RGB camera is to improve geolocation of HSI data collected by the main payload, versus geolocation solely derived from spacecraft bus attitude estimation.

High-quality geolocation of HSI data has been identified as important for scientific uses of the HSI data products. Camera imagery will be used to identify and geolocate ground features with known ground-truth locations to improve attitude estimation. The RGB camera will also be used to provide visible light imagery of the ROI for external communication purposes, while possible secondary uses of the RGB camera imagery for scientific purposes is still being studied.

To this end, a variety of potential focal lengths were considered. Given a fixed image sensor size and resolution, focal length determines the FoV of the RGB imager, as well as the ground resolution. A longer focal length results in a smaller FoV and improved ground resolution, but at the cost of requiring the ocean area being studied to be closer to a ground feature, suitable for use as a ground-truth location, such that both appear in a single imager frame. Finally, a focal length of 4.4 mm was settled on, with a Horizontal and Vertical FoV at full angle of 44.3 and 34 respectively, to provide a good compromise between FoV and feature resolution.

## 5 Spacecraft Platform

The AEROS spacecraft is based on the OpenSat 3U platform from OpenCosmos. The spacecraft has been designed for a 3-year lifetime at SSO, guaranteeing all functions to support payload operations.

The main subsystems, including Electrical Power System (EPS), On-Board Data Handling (OBDH), Telemetry & TeleCommand (TTC) and Attitude Determination and Control (ADCS), are all flight proven (TRL 9), minimising mission risk. The current spacecraft configuration can be seen in Figure 7.

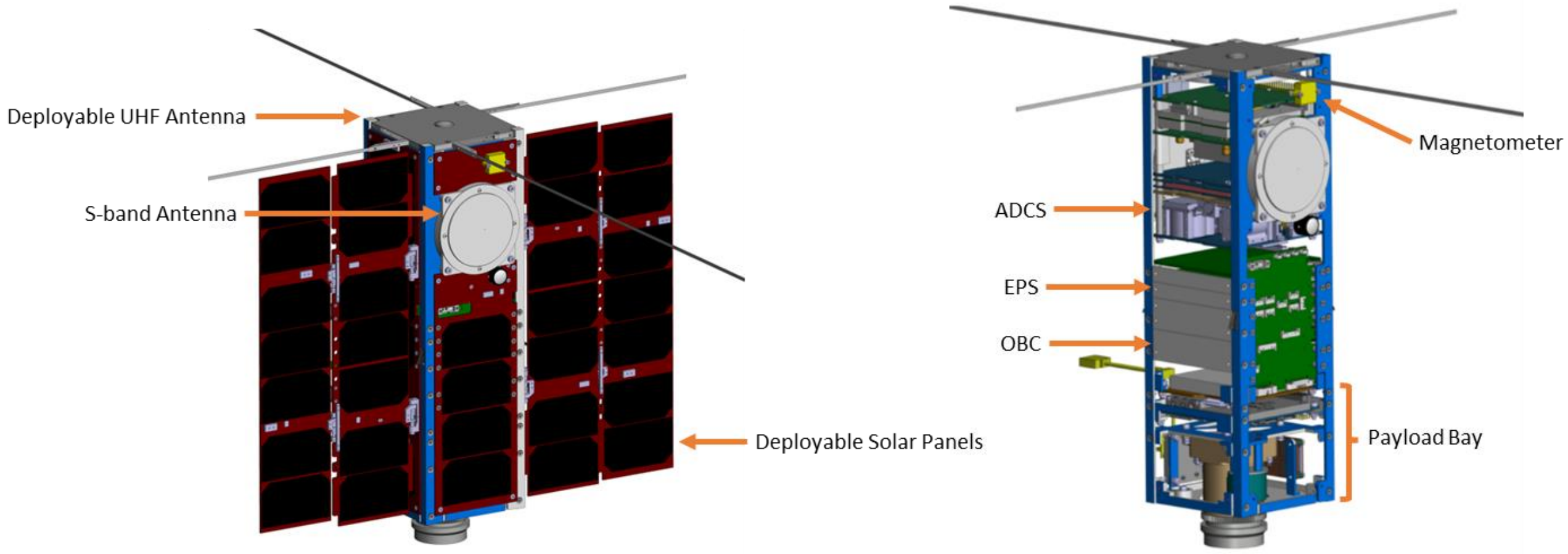

Figure 7. Satellite Overall View (Left) and Electronics Integration View (Right). Credit: CEiiA

On top of the OpenCosmos solution, some developments will be implemented, namely the structures and the OBC software (an adaptation of the RTEMS software), guaranteeing a reliable optimised system. This results in a 5.5 kg 3U (119 mm x 119 mm x 385 mm) spacecraft, with 45.6 Wh battery capacity, a S-Band ground link (Downlink: 512 kbit/s; Uplink: 256 kbit/s) and a pointing capability of ±2°.

### 5.1 Structures

The spacecraft structures, based on a Hard Anodized Aluminium solution being developed by CEiiA, have been designed to minimise mass, guaranteeing the maximum volume for the payload bay (Figure 8). The custom structural solution in development relies in Computer Numerical Control machining, which allows high flexibility in the design stage and increased precision during manufacturing and assembly. Additionally, several Finite Element Method analyses are being performed, based in static and dynamic loads necessary to comply with the flight envelopes imposed by different launchers. This structural analysis also includes the optimization of vibrations response of the complete satellite, ensuring that maximum G Root-Mean-Square are not exceeded, and the natural vibration frequencies of the payload are not superimposed to the overall satellite response, preventing vibratory resonance events.

This structural approach will allow the optimization of the payload area, overall structural mass and volume, as well as to ensure the structural integrity of the satellite and consequently increase competitiveness of the final solution.

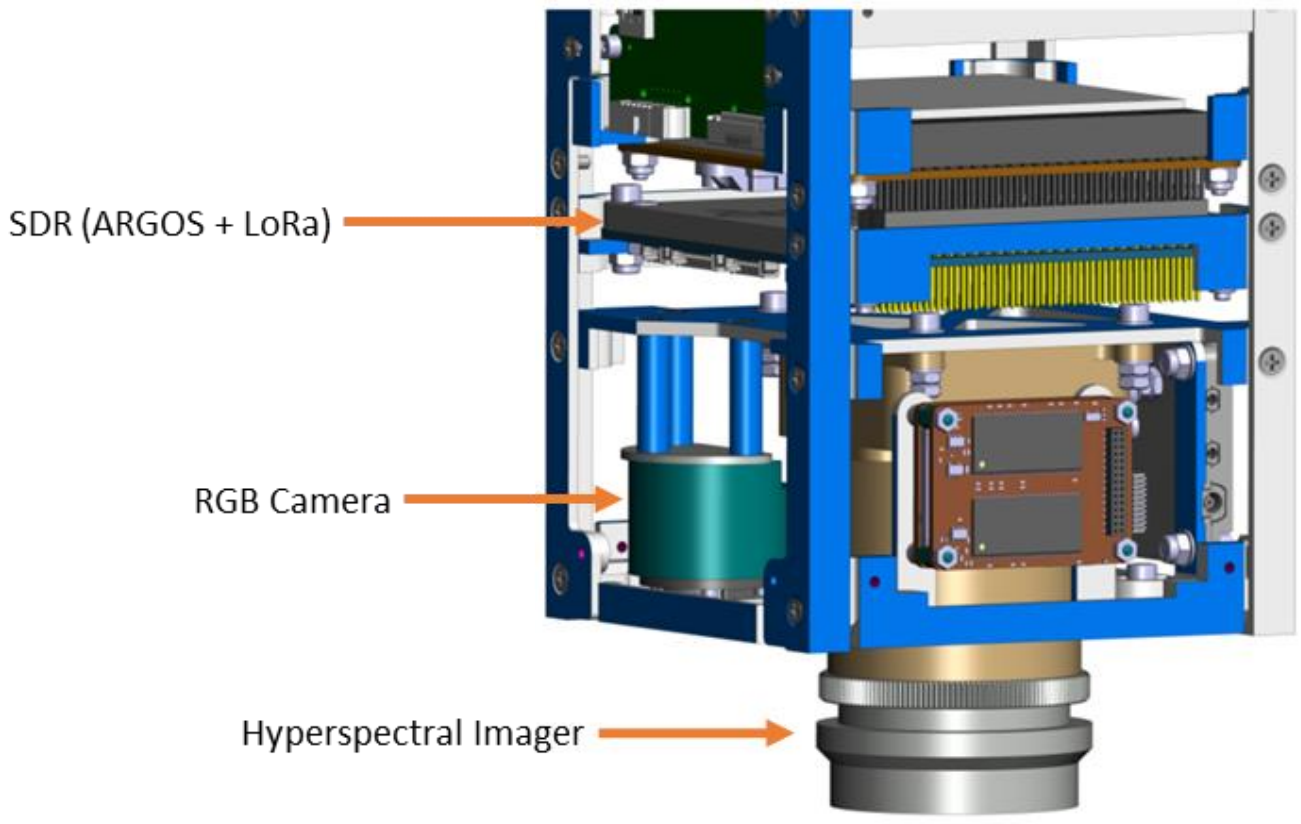

Figure 8. AEROS Spacecraft Payload Bay Arrangement. Credit: CEiiA

### 5.2 Software

The adapted RTEMS, aka RTEMS Improvement, is a space pre-qualified real time operating system developed by EDISOFT and used in more than 40 ESA (European Space Agency) missions (including Galileo FOC, Solar Orbiter, and Sentinel 2) and included in four missions of South Korea. RTEMS is a Real-Time Operating System for Multiprocessor Systems. It is open source and aims to be competitive with closed source commercial products. It has been developed to support applications with the most inflexible timeframe requirements, making it possible for the user to develop hard real-time systems.

The AEROS spacecraft OBC software will run in a single computer and be used to control the spacecraft itself (having peripherals with device drivers), and to communicate with the implemented payloads. The integration can be performed with the devices apart and using Transmission Control Protocol (TCP)- Internet Protocol (IP) sockets and translators between TCP-IP protocol and the devices protocol (UART, CAN, etc). The OBC use an AM3358BZCZA100 processor, with the characteristics shown in Table 2.

Table 2. AEROS OBC Specifications

| | |
|---|---|
| **CPU** | ARM Cortex A8 32 Bit, 1 GHz |
| **Cache Memory** | 64 KB On-Chip Memory (Shared L3 RAM) |
| **Volatile Memory** | 256 MB DDR2 RAM:<br>• General-Purpose Memory Controller with BCH code to Support ECC<br>• Error Locator Module |

| | | |
|---|---|---|
| **Real-Time Unit** | Programmable Real-Time Unit and Industrial Communication Subsystem:<br>• 1 x UART (up to 12 Mbps)<br>• 1 x eCAP module<br>• 2 x MII ethernet ports<br>• 1 x MDIO | |
| **Real-Time Clock** | Internal 32.768 kHz oscillator | |
| **System Peripherals** | *Used in AEROS* | |
| | CAN Interface | 2 x CAN ports (version 2 Part A and B) |
| | Serial Interfaces | 6 x UART (support IrDA, CIR Modes, RTS and CTS flow control) |
| | Primary Storage | 32 GB EMMC |
| | *Other available* | |
| | SPI Interfaces | 2 x Master and Slave McSPI |
| | I2C Interfaces | 3 x I2C Master and Slave |
| | I/O | 4 x banks of GPIO with 32 GPIO pins per bank |

# 6 DATA CENTRE

## 6.1 Data Storage and Data Display

The DAC, being developed by CEiiA, can be divided into three main parts, Back-end, Extract, Transform, Load (ETL) and Front-end, as seen in Figure 9.

The DAC begins with acquiring data from several sources, being capable to integrate and manage different types and layers of data. After obtaining the data, the DAC uses the ETL process which considers the activities of data extraction, data parallelization (data validation and transformation), and data loading. Once the data has been processed and loaded into a target database, it will be possible to view tailored-to-use remote sensing products being developed as part of the AEROS mission, in the front-end web app, which uses a modular approach with design system concepts like atomic design.

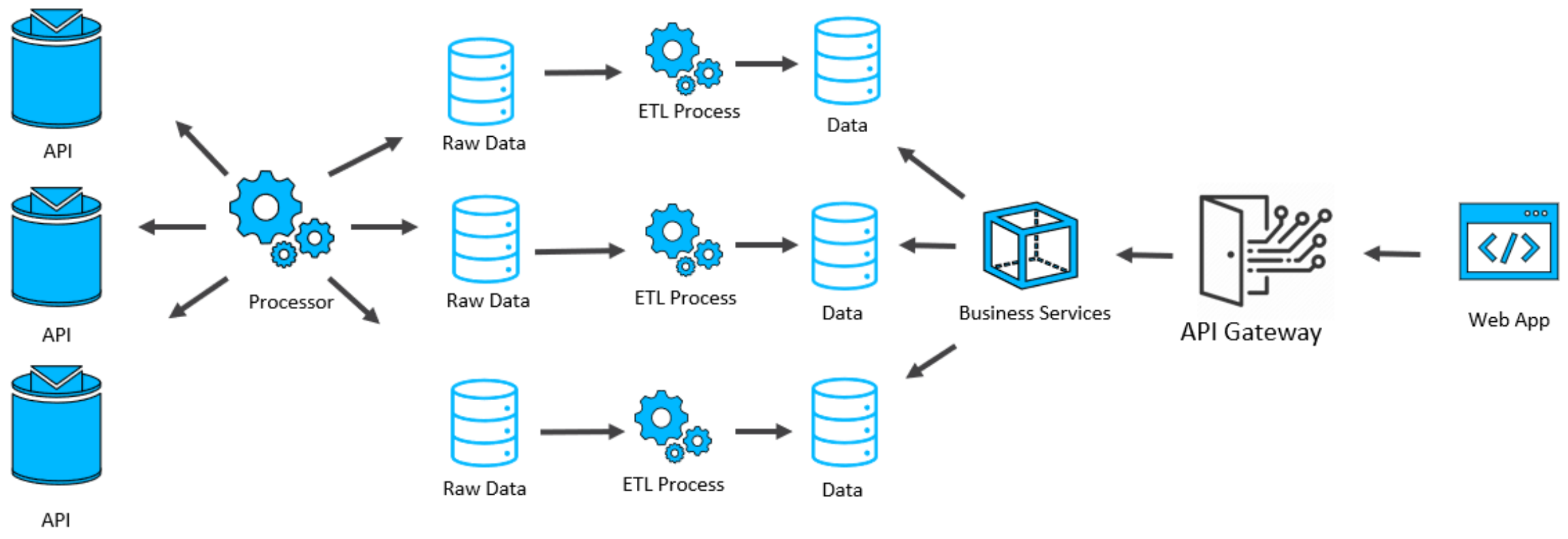

Figure 9. Conceptual Architecture. Credit: CEiiA

## 6.2 Data Applications

As part of the DAC, and considering the AEROS use cases and objectives (Section 2 and Section 3), algorithms to derive Inherent Optical Properties (IOPs) of several optically active constituents including Chlorophyll-a (Chl-a) and CDOM, and models to estimate SSS, are being studied and developed.

### 6.2.1 *Ocean Colour*

Ocean colour images provides data on the concentrations of chlorophyll (Chl), whose measurements provide the basis for global and regional ecosystems studies, allowing to quantify and monitor the spatial and temporal variability of the phytoplankton biomass [27]. Phytoplankton are marine plants that get their energy via photosynthesis, meaning they contain substantial amounts of Chl. Depending on the type of concentration of phytoplankton in a given region, ocean colour varies in shades of blue and green.

Remote sensing of ocean colour allows then to measure phytoplankton and their impact on the Earth system. Information about optical active constituents present in water such as Chl-a and CDOM, and optical properties such as absorption and backscattering can be obtained from its spectral reflectance spectrum, and the accuracy of this information is dependent of the quality of the estimated remote sensing reflectance ($R_{rs}(\lambda)$) [21], [27],[30]. The remote sensing reflectance, applicable to the current ocean colour sensors, is represented by [30]:

$$R_{rs}(\lambda) = \Re \frac{f(\lambda)}{Q(\lambda)} \frac{b_b(\lambda)}{a(\lambda)+b(\lambda)}, \quad (1)$$

where $a(\lambda)$ is the spectral total absorption coefficient, $b(\lambda)$ the spectral total backscattering coefficient, $\Re$ is a factor that accounts for the reflection and refraction at air-water interface, and $f/Q$ accounts for the bidirectional nature of the reflectance.

Satellite ocean colour data can then provide relevant information for coastal management, fisheries and aquaculture management, ecosystem assessment and biogeochemistry, helping to contextualize behavioural shifts within our changing climate [22], [31].

Algorithms to measure concentration of Chl-a have been develop using an empirical relationship derived from *in situ* measurements of Chl-a and remote sensing reflectance [32]. The abundance-based algorithms, based on the general observation that in the open ocean a change in Chl is associated with a change in phytoplankton composition or size structure, have also been widely used [30].

Reference [33] evaluated 65 empirical ocean colour chlorophyll algorithms for 25 satellite instruments. The results showed that bands at 412 nm for the estimation of chlorophyll present good results. It also showed that the bands at 443 nm, 490 nm, 510 nm and 555 nm can be used for chlorophyll retrieval, with the 673 nm showing a secondary peak of chlorophyll. This is also supported by [31], where it is proposed that satellites that want to detect ocean colour, should have a minimum number of bands. The role of the various bands is presented in Table 3.

Table 3. Ocean Colour Information and the Role of the Bands as Seen in [31]

| Band Wavelengths (nm) | Proxy |
|---|---|
| 413 | Discriminate CDOM |
| 443, 490, 510, 560 | Chlorophyll retrieval from blue-green ratio algorithms |
| 560, 620, 665 and others | Retrieve water content in turbid Case 2 waters using new red-green algorithms |
| 665, 681, 709 and others | Chlorophyll retrieval using fluorescence peak |
| 779, 870 | Atmospheric correction |

### 6.2.2 *SEA SURFACE SALINITY*

SSS is one of the most important parameters for predicting and understanding ocean currents (related with water cycle dynamics) and climate, and the best indicator of freshwater exchanges. SSS can be described as the salt concentration in the upper centimetre of the ocean surface [34], measured in Practical Salinity Units (PSU).

Salinity is linked to the mass and heat balances between the atmosphere and the ocean through evaporation, precipitation, and river runoff [35]. This exchange of freshwater between atmosphere and ocean leaves a strong imprint on ocean salinity, since precipitation and runoff bring freshwater into the ocean, diluting and, therefore, decreasing the salt content of the surface water. Evaporation from the ocean's surface on the other hand, leaves salt behind, increasing its salinity [35]. This variations affect biological and physical processes in the ocean, which can lead to dangerous long-range water cycle and ocean circulations effects, affecting regional climates and marine life [36].

Mapping SSS has also supported the studies that point to climate warming as one of the responsible for the increase of extreme global water cycle changes [35]. Moreover, salinity plays a part in mitigation climate change, since the amount of gases that can be dissolved in seawater depends on salinity and temperature, meaning that an increase in these parameters results in a decrease in the amount of $CO_2$ that can be absorbed by the oceans [37].

Observing and mapping the variability of ocean salinity is, therefore, valuable to understand and predict ocean dynamics and extreme climate events, such as floods and droughts, that represent a risk to the world's food security, to human health and many other ecosystems, and to the global economy.

The use of satellite remote sensing in monitoring SSS combined with the information obtained from *in situ* platforms, such as vessels and buoys, has been innovator thanks to the possibility of creating maps of the global ocean with a higher-temporal resolution, making SSS observations more accurate.

Following ocean colour detection methods that use the remote sensing reflectance variable, in [4], a model for estimating SSS using $R_{rs}$ from the Geostationary Ocean Color Imager (GOCI) abord the Communication, Ocean and Meteorological Satellite, showed a good estimation of SSS making it a relevant case for the AEROS mission.

To verify if SSS and $R_{rs}(\lambda)$ values had a close relationship, a multiple linear regression was applied to a single band, band ratios and different combinations. The bands wavelengths were: 412 nm, 443 nm, 490 nm, 555 nm, 660 nm, and 680 nm. Reference [4] ended with a model that was able to identify which GOCI bands demonstrated the strongest correlation between SSS and $\mathrm{R_{rs}}(\lambda)$ with low errors and high coefficient of determination values. This approached was able to demonstrate that SSS can be extracted from measured spectral reflectance on the visible spectral bands.

Reference [38] also showed that estimated of SSS can be expressed directly as a function of remotely sensed ocean colour bands using Moderate Resolution Imaging Spectroradiometer (MODIS)-Aqua bands. The results indicated that the remote sensing reflectance at 488 nm was positively associated with salinity.

As AEROS proposes to estimate ocean salinity, a model is being developed to automatically estimate SSS from multispectral images. This information will then be available in the DAC of AEROS ready to be visualized by the end user.

#### *6.2.3 CDOM*

There are other substances in the water that are capable of absorbing light. These are usually composed of organic carbon, being usually refer to as CDOM.

While there is a big number of ocean colour algorithms that estimate Chl-a concentrations as seen above, there are few algorithms to estimate CDOM and salinity using the remote sensing reflectance variable [39]. With several studies having found that detritus and CDOM concentrations are good tracers of salinity [38], [40], [41], the focus in AEROS is on the research of [39], that using simple band ratio approach, developed an algorithms able to retrieve CDOM absorption and salinity values from the derived CDOM absorption coefficients. The results concluded that the remote sensing reflectance at 665 nm and 490 nm were the best combination for CDOM absorption model.

## 7 SUMMARY

As climate change becomes more evident, the need for protecting the oceans becomes essential. As remote sensing technologies evolve, small satellites prove to be an efficient and cost-effective tool for the rapid need of monitoring the oceans.

Ocean colour measurements allow to estimate inherent optical properties (IOPs) of ocean characteristics such as Chl-a, phytoplankton and CDOM. This data is crucial to obtain SSS maps and can give information about water quality parameters which are important for maintaining the MPAs.

The proposed objectives of the AEROS mission are critical to understand the overall marine ecosystem functioning, predicting, and managing MPAs, helping with aquaculture management and commercial sectors such as fisheries and tourism, important on a socio-economic level. In the end, the AEROS Data Analysis Centre will gather new information for future customers and stakeholders, providing new remote sensing products.

With the main AEROS use cases established, mission objectives and system requirements have been set, with a successful System Requirement Review held. Among those requirements are the main AEROS payload targeted frequencies, to be used to detect these ocean phenomena (Table 4).

Table 4. AEROS Target Bands for the Ocean Parameters

| Proxy | Wavelength Centre (nm) |
|---|---|
| Ocean Colour | 475, 495, 545, 555, 625, 865 |
| Sea Surface Salinity | 485, 545, 555, 625, 685, 865 |
| CDOM | 655 |

From those requirements, the procurement of the main platform and payload components has been completed, which were a priority given the associated long lead times (in particular with the world's shortage of electronic components).

While the lead time is being fulfilled, the remaining components, including structures, OBC software, HSI and SDR payloads and the DAC and needed algorithms are being developed.

All these developments, aimed at guaranteeing the mission success (in particular the retrieval of high confidence data), will be reviewed by the middle of the current year in a mission major Design Review. This planning is in line with the target launch by middle of 2023, give the available slots from Falcon 9, and a mission demonstration by the end of the year.

The launch and operation of the AEROS pathfinder, in addition to the mission goals, will impactfully promote the advancement of Portuguese scientific and technological knowledge in the Space industry by joining the list of small satellites used for EO, ensuring Portugal's rise as a robust space-faring nation.

With a successful operation on-orbit of AEROS, a future constellation can be prepared, based on ten satellites at different inclinations. This constellation will reduce the revisit time down to less than one hour, achieving this way a temporal resolution that will provide the market with precise and timely insights on the Atlantic, as well as in the rest of the globe, promoting smart, autonomous, and sustainable monitoring of the ocean.

## ACKNOWLEDGMENTS

The AEROS project (Nr. 045911) leading to this work is led by a Portuguese consortium and co-financed by the ERDF - European Regional Development Fund through the Competitiveness and Internationalisation - COMPETE 2020, PORTUGAL 2020, LISBOA 2020 and by the Portuguese Foundation for Science and Technology - FCT under the International Partnership programme MIT Portugal. The Azorean partners are co-financed by AÇORES 2020, project number ACORES-01-0145-FEDER-000131, by FCT under the project UIDB/05634/2020, DRCT the Regional Government of the Azores through the initiative to support the Research Centres of the University of the Azores and through the project M1.1.A/REEQ.CIENTÍFICO UI&D/2021/010. J. Fontes was also co-financed by AÇORES 2020, through the Fund 01-0145-FEDER-000140. The authors also acknowledge AGI, an Ansys company and its Educational Alliance Program, for donating its Systems Tool Kit (STK) software, which was used to conduct orbital analysis by the MIT team for the AEROS mission.